\documentclass[11pt,twoside]{article}

\usepackage{asp2006}
\usepackage{epsf}
\usepackage{lscape}

\begin{document}
\title{Stokes Profiles at the Narrow Magnetic Lanes of Sun Spots}   
\author{Gordon A. MacDonald, Kemal A. Yassin and Debi Prasad Choudhary}   
\affil{Department of Physics and Astronomy,  
California State University Northridge,  
18111 Nordhoff Street  Northridge, CA 91330-8229}    

\begin{abstract} 
It has been previously observed that narrow lanes of transverse and longitudinal
magnetic field with opposite polarity are the site of large solar flares \citep{zw1993}. We
performed a comprehensive examination of the stokes asymmetries of active
region NOAA 10930. The active region was observed just before, during and
after an X-class flare, which occurred during December 13, 2006 from 02:20 to
06:18 UT. We observe a static fibril interacting with a rotating penumbra of
opposite polarity in the hours prior to the flare. Above the fibril were several
small sites of hot gas in the chromosphere. During and after the flare, the fibril
and its corresponding flow and profiles were much less pronounced. We present
a full analysis of magnetic and plasma properties of this active region.
\end{abstract}


\section{Introduction} Our sun is an average star in the Milky Way Galaxy.
Understanding the fundamental processes in the sun system is key to understanding
phenomena in the earth and solar system. One of these are solar magnetic
phenomena known as sun spots, which are somehow connected violent explosions
in the solar atmosphere known as solar flares. Using data from the Solar Optical
Telescope on board the Hinode Mission, we observed a solar flare which took place
in AR NOAA 10930 on December 13, 2006. The Stokes IQUV parameters of the FeI
6301.5Å line were analyzed for Doppler shift, Stokes V zero crossing, and Stokes
V amplitude and area asymmetry using the techniques developed by \citet{dc2009}.
The result of this study is a comprehensive analysis of the
spectropolarimetric data of AR NOAA 10930, with attention directed in particular
to narrow lanes of magnetic field between the two spots in this region. Four
regions were selected in this area, and their stokes profiles are presented.

\section{Data Reduction} For a detailed description of our analysis, please see \citet{dc2009}. In this section, the construction of the dopplergrams, Stokes V zero-crossing maps and amplitude and area asymmetry in the Stokes V parameter are presented. The Stokes \(I, Q, U\) and \(V\) profiles were normalized by taking the average \(I\) profile of 100 pixels in a region of quiet continuum. From this average \(I\) profile, the continuum intensity was obtained in order to normalize the remaining three profiles, as well as the Fe I absorption line separation which gave the dispersion. A transformation was then performed which resulted in the Stokes profiles plotted at a function of wavelength.

\par A Doppler transform was performed on a map containing the Stokes I line centers in the FOV. The distribution of the doppler velocities are normal and thus any pixels giving a speed more than three standard deviations from the mean were rejected. Most of these outliers were due to anamolous profiles. An identical treatment was applied to the zero-crossing maps obtained from the Stokes V profile.

\par The normalized Stokes V profiles were used in determining their amplitude and area asymetry. The asymmetry is obtained using the following definitions \citep{ss1984}
\begin{equation}
\delta a = \frac{\left| a_b \right| - \left| a_r \right|}{\left| a_b \right| + \left| a_r \right|}
\end{equation}

\begin{equation}
\delta A = \frac{\left| A_b \right| - \left| A_r \right|}{\left| A_b \right| + \left| A_r \right|}
\end{equation}
\\where \(a\) represents amplitude and \(A\) represents area of a Stokes V lobe and the \(\delta\) represents the asymmetry in either of the two.

\section{Observatons} The size of the area containing the narrow magnetic lanes increases dramatically during the flare and decreases in size after the flare. This is seen in the dopplergrams and zero-crossing maps in \textit{figure 1}. Very intricate structures evolve and change dramatically during the period of observations. Areas of strong flow developed during the flare, then disappeared after. LOS flow speeds were as high as 2.5 km/s in the active region.
\par Highly disordered patterns of islands and gulfs are seen in the asymmetry figures before the flare. This pattern becomes more orderly during and after the flare.

\begin{figure}[!ht]
\plotone{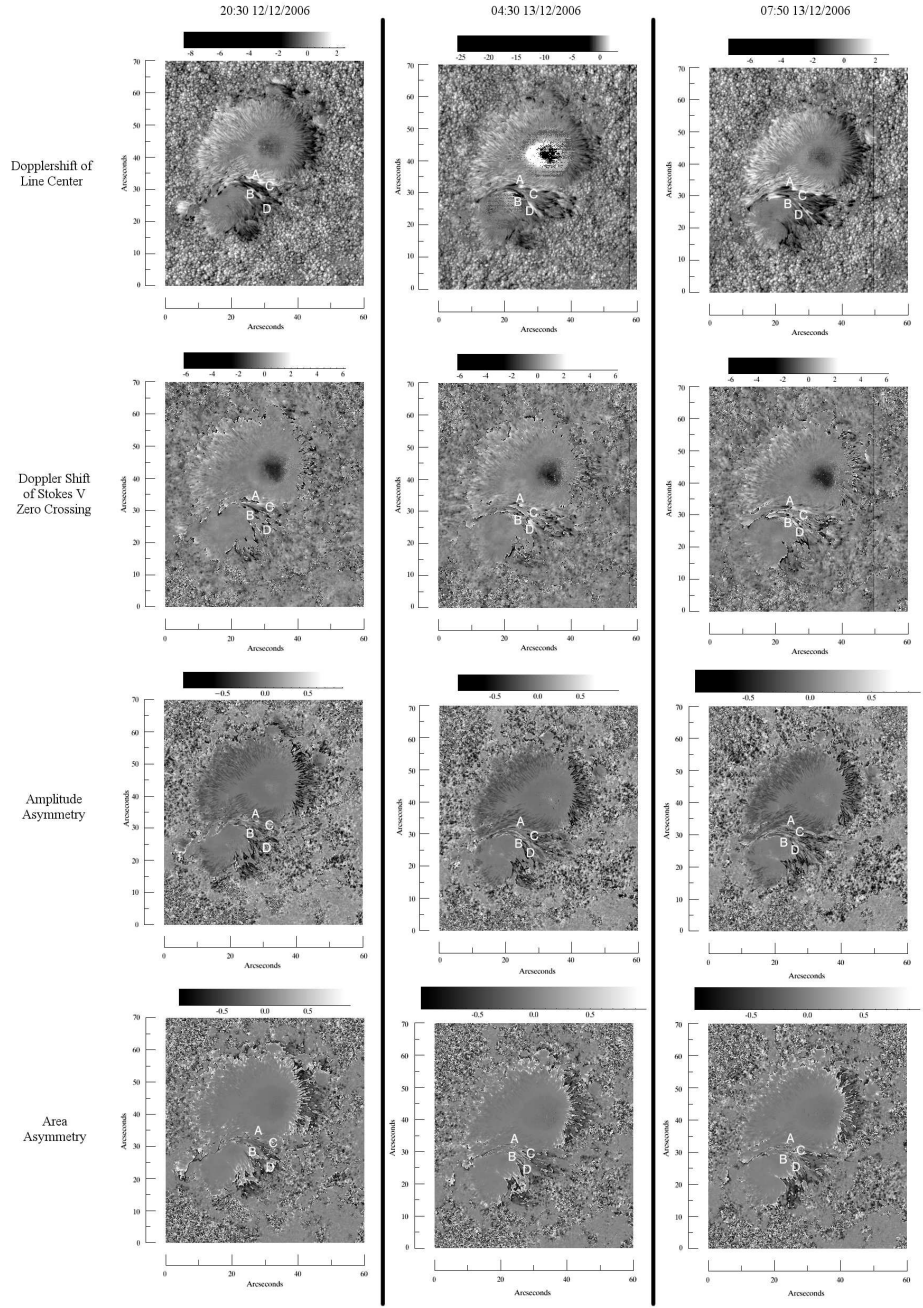}
\caption{Time evolution of Doppler flow, magnetic flow, and amplitude and area asymmetry in the Stokes V parameter. Each of the three columns correspond to before, during and after the flare. Regions A through D are specified on each sub-figure. Stokes profiles at regions A through D are seen in Figure 2.}
\end{figure}

\begin{figure}[!ht]
\plotone{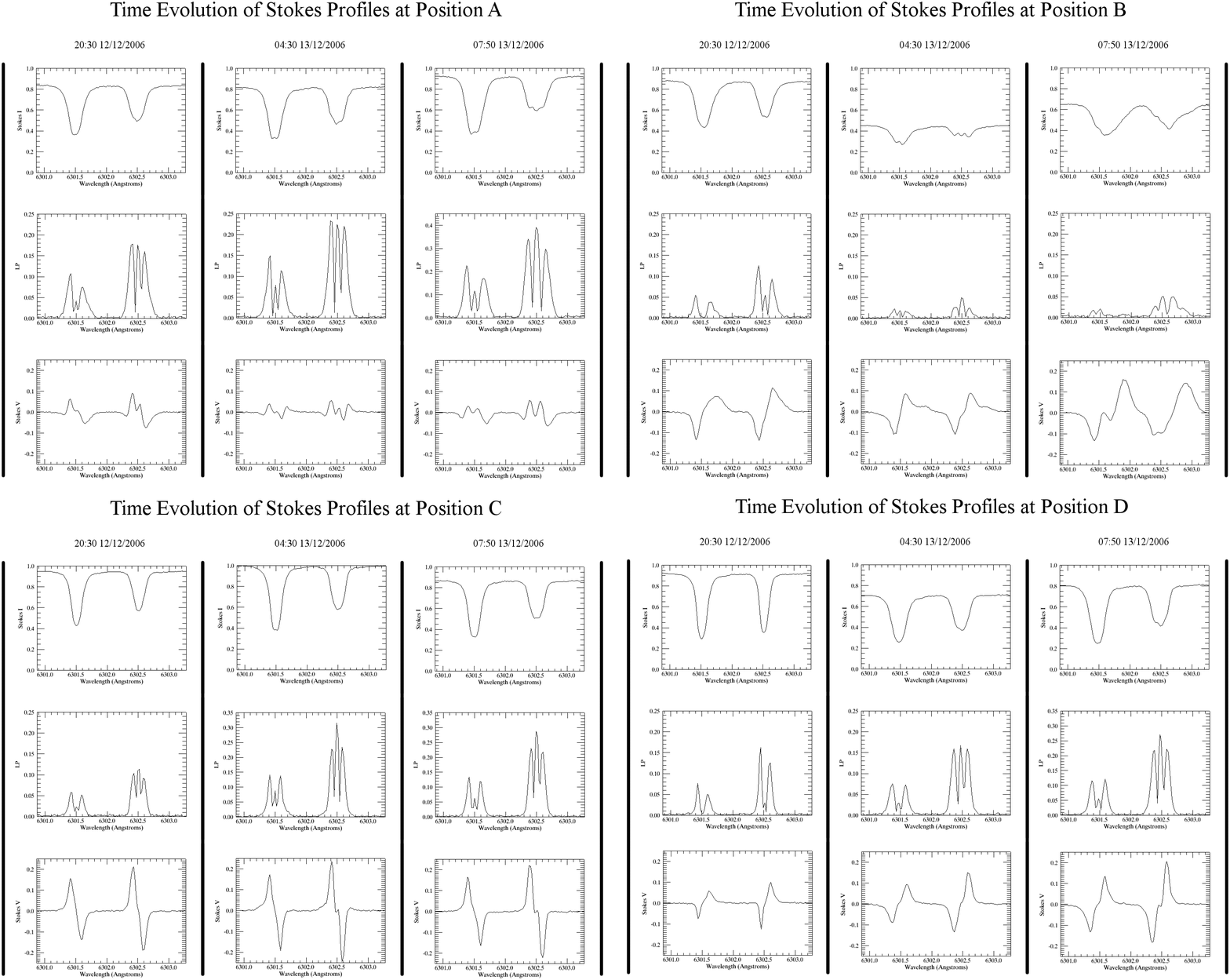}
\caption{Stokes profiles for both FeI 6301.5Å and 6302.5Å at positions A through D. The profiles shown are Stokes I and V and the total linear polarization magnitude \( LP=\sqrt{Q^2 + U^2} \) . The columns in each of the three sub-figures correspond to before during and after the flare. Many of the profiles are anomalous, most notably, Stokes V.}
\end{figure}

\acknowledgements This work was supported by NSF grant ATM-0548260 at California
State University Northridge. The data was obtained using Hinode Solar Optical
Telescope. Hinode is a Japanese mission developed and launched by ISAS/JAXA, with
NAOJ as domestic partner and NASA and STFC (UK) as international partners. It is
operated by these agencies in co-operation with ESA and NSC (Norway).


\end{document}